\documentstyle[12pt,epsfig]{article}
\begin{document}

\begin{flushright}
{\bf hep-ph/0102304} \\
{\bf LMU-01-03} 
\end{flushright}

\vspace{0.5cm}

\begin{center}
{\Large\bf Nearly Bi-Maximal Neutrino Mixing, Muon g-2 Anomaly 
and Lepton-Flavor-Violating Processes}
\end{center}

\vspace{0.1cm}

\begin{center}
{\bf Zhi-zhong Xing
\footnote{Electronic address: xingzz@mail.ihep.ac.cn}} \\ 
{\it Institute of High Energy Physics, P.O. Box 918,
Beijing 100039, China} 
\end{center}

\vspace{2.5cm}

\begin{abstract}
We interpret the newly observed muon g-2 anomaly
in the framework of a leptonic Higgs doublet model with nearly
degenerate neutrino masses and nearly bi-maximal neutrino mixing.
Useful constraints are obtained on the rates of
lepton-flavor-violating rare decays $\tau \rightarrow \mu \gamma$, 
$\mu \rightarrow e \gamma$ and $\tau \rightarrow e \gamma$ as well
as the $\mu$-$e$ conversion ratio $R_{\mu e}$. We find that 
$\Gamma (\mu \rightarrow e \gamma)$, $\Gamma (\tau \rightarrow e \gamma)$ and
$R_{\mu e}$ depend crucially on possible non-zero but samll values of
the neutrino mixing matrix element $V_{e3}$,
and they are also sensitive to the Dirac-type CP-violating phase.
In particular, we show that 
$\Gamma (\tau \rightarrow \mu \gamma)/m^5_\tau$, 
$\Gamma (\mu \rightarrow e \gamma)/m^5_\mu$ and 
$\Gamma (\tau \rightarrow e \gamma)/m^5_\tau$
are approximately in the ratio $1: 2|V_{e3}|^2: 2|V_{e3}|^2$ if
$|V_{e3}|$ is much larger than ${\cal O}(10^{-2})$, and
in the ratio 
$2 (\Delta m^2_{\rm atm})^2: (\Delta m^2_{\rm sun})^2:(\Delta m^2_{\rm sun})^2$
if $|V_{e3}|$ is much lower than ${\cal O}(10^{-3})$, where
$\Delta m^2_{\rm atm}$ and $\Delta m^2_{\rm sun}$ are the corresponding
mass-squared differences of atmospheric and solar neutrino oscillations.
\end{abstract}


\newpage

Recently the Muon g-2 Collaboration has reported a precise measurement
of the muon anomalous magnetic moment \cite{EX},
\begin{equation}
a_\mu ({\rm exp}) \; =\; \frac{g^{~}_\mu - 2}{2} \; =\;
11659202(16) \times 10^{-10} \; ,
\end{equation}
which deviates from the standard-model (SM) prediction by 2.6$\sigma$:
\begin{equation}
\Delta a_\mu \; \equiv \; a_\mu ({\rm exp}) - a_\mu ({\rm SM}) \; =\;
43(16) \times 10^{-10} \; .
\end{equation}
To interpret the nonvanishing and positive value of $\Delta a_\mu$, a
lot of scenarios of new physics have been proposed \cite{Mac}--\cite{Ma}. 
In Ref. \cite{Ma}, Ma and Raidal emphasized that the new physics responsible 
for $\Delta a_\mu$ and other lepton-flavor-violating processes (e.g.,
$\mu \rightarrow e\gamma$) might be related to that responsible for
non-zero neutrino masses indicated by the evidence of neutrino oscillations.
They illustrated this important point in the framework of a leptonic Higgs doublet 
model \cite{Ma2} with nearly degenerate neutrino masses and bi-maximal 
lepton flavor mixing, and found that 
$\Gamma (\mu \rightarrow e \gamma)/m^5_\mu$ and
$\Gamma (\tau \rightarrow \mu \gamma)/m^5_\tau$ are proportional 
respectively to $(\Delta m^2_{\rm sun})^2$ and $(\Delta m^2_{\rm atm})^2$,
where $\Delta m^2_{\rm sun}$ and $\Delta m^2_{\rm atm}$ correspond to
the mass-squared differences of solar and atmospheric neutrino oscillations.

In this paper we take the same Higgs doublet model to examine whether
the interesting relations such as 
$\Gamma (\mu \rightarrow e \gamma)/m^5_\mu \propto (\Delta m^2_{\rm sun})^2$
keep unchanged, if the bi-maximal neutrino mixing pattern ($|V_{e3}|=0$)
is replaced by the {\it nearly} bi-maximal neutrino mixing pattern 
($|V_{e3}| \neq 0$). The latter is more realistic to fit various neutrino
oscillation data, and it can accommodate large CP violation which might be
observable in the future long-baseline neutrino experiments.
While $\Delta a_\mu$ and $\Gamma (\tau \rightarrow \mu \gamma)$
are insensitive to non-zero but small $|V_{e3}|$, we find that 
$\Gamma (\mu \rightarrow e \gamma)$,
$\Gamma (\tau \rightarrow e \gamma)$ and the $\mu$-$e$ conversion ratio
$R_{\mu e}$ {\it do} depend crucially upon the magnitude of $|V_{e3}|$.
The simple relations  
$\Gamma (\mu \rightarrow e \gamma)/m^5_\mu \propto (\Delta m^2_{\rm sun})^2$
and $\Gamma (\tau \rightarrow e \gamma)/m^5_\tau \propto (\Delta m^2_{\rm sun})^2$
may still hold, if $|V_{e3}|$ is much lower than ${\cal O}(10^{-3})$. If
$|V_{e3}|$ is much larger than ${\cal O}(10^{-2})$, however, the rates of
$\mu \rightarrow e \gamma$, $\tau \rightarrow \mu \gamma$ and
$\tau \rightarrow e \gamma$ will all be proportional to $(\Delta m^2_{\rm atm})^2$.
Our new results illustrate that an accurate determination of 
$|V_{e3}|$ from neutrino oscillations is desirable, in order to predict or
constrain the rates of lepton-flavor-violating rare decays and to fix
the Dirac-type CP-violating phase in the lepton mixing matrix.

Let us concentrate on the leptonic Higgs doublet model proposed in 
Ref. \cite{Ma2}. The neutrino mass matrix $M_\nu$, defined in the flavor
basis where the charged lepton mass matrix is diagonal, can be derived 
from the following Lagrangian through a simple seesaw mechanism:
\begin{equation}
\frac{1}{2} M_i N^2_{i \rm R} ~ + ~ h_{ij} \overline{N}_{i \rm R} 
\left ( \nu^{~}_j \eta^0 - l_{j \rm L} \eta^+ \right ) ~ + ~ {\rm h.c.} \; .
\end{equation}
The typical Feynman diagram giving rise to $\Delta a_\mu$ and
$l_i \rightarrow l_j \gamma$ is shown in Fig. 1. Following Ma and 
Raidal \cite{Ma}, we assume that the masses of three $N_{\rm R}$'s are equal, 
and the coupling matrix $(h_{ij})$ takes the form
\begin{equation}
(h_{ij}) \; =\; 2 \left ( \matrix{
h_1	& 0	& 0 \cr
0	& h_2	& 0 \cr
0	& 0	& h_3 \cr} \right ) V^{\rm T} \; ,
\end{equation}
where $h_i$ (for $i=1,2,3$) are the eigenvalues of $(h_{ij})$, and $V$ is
the neutrino mixing matrix in the chosen flavor basis. The neutrino mass
eigenvalues $m_i$ can then be calculated by using the seesaw relation
$M_\nu = (h_{ij})^{\rm T}(h_{ij})/M$, where $M$ is the heavy mass scale 
common for three $N_{\rm R}$'s. Explicitly one obtains 
$m_i = 4 h^2_i \langle \eta^0 \rangle^2/M$. Different 
from Ref. \cite{Ma}, the neutrino mixing matrix $V$ is taken to be
nearly bi-maximal in this paper:
\begin{equation}
V \; =\; \left ( \matrix{
\frac{c}{\sqrt{2}}	& \frac{c}{\sqrt{2}}	& -is \cr\cr
-\frac{A}{2}	& \frac{A^*}{2}	& \frac{c}{\sqrt{2}} \cr\cr
\frac{A^*}{2}	& -\frac{A}{2}	& \frac{c}{\sqrt{2}} \cr} \right ) \; ,
\end{equation}
where $s \equiv \sin \theta$, $c\equiv \cos\theta$, and $A = 1 + is$ \cite{Xing}.
The mixing angle $\theta$ measures a slight coupling between solar and
atmospheric neutrino oscillations. The current experimental constraint on 
the magnitude of $s$ is $s \leq 0.2$, obtained from CHOOZ and Palo Verde reactor 
experiments of neutrino oscillations \cite{CHOOZ}. Note that only the
Dirac-type CP-violating phase (of a special but instructive value $90^\circ$) 
has been taken into account in $V$, since the Majorana-type phases have no effect 
on the rare processes to be discussed subsequently. 
\begin{figure}[t]
\vspace{-3.2cm}
\epsfig{file=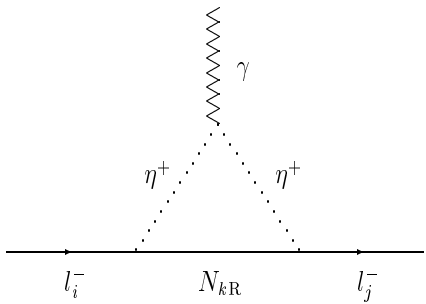,bbllx=-3cm,bblly=6cm,bburx=16cm,bbury=28cm,%
width=15cm,height=20cm,angle=0,clip=}
\vspace{-12.5cm}
\caption{The Feynman diagram giving rise to $\Delta a_\mu$ and 
$l_i \rightarrow l_j \gamma$ in the leptonic Higgs doublet model,
where the photon can be attached to any charged line.}
\end{figure}

For simplicity, we follow Ref. \cite{Ma} to make two more assumptions: 
(a) the masses of three light neutrinos are nearly degenerate, i.e., 
$m_3 \approx m_2 \approx m_1 \equiv m$ or equivalently
$h_3 \approx h_2 \approx h_1 \equiv h$; and (b) the univeral mass of 
$N_{\rm R}$'s is identical to the mass of $\eta$. 
While assumption (b) helps to make the value of $\Delta a_\mu$ as large
as possible within the model under consideration, 
assumption (a) is necessary for the suppression of 
$\Gamma (\tau \rightarrow \mu \gamma)$ relative to $\Delta a_\mu$.
Calculating Fig. 1, one obtains
\begin{equation}
\Delta a_\mu \; =\; \sum_i \frac{|h_{i\mu}|^2}{192 \pi^2} 
\cdot \frac{m^2_\mu}{m^2_\eta} \; .
\end{equation}
With the help of Eqs. (4) and (5) as well as the assumptions made above, 
we can simplify Eq. (6) and arrive at 
\begin{equation}
m_\eta \; \approx \; \frac{m_\mu}{2 \sqrt{3\pi \Delta a_\mu}} 
\sqrt{\alpha^{~}_h} \; ,
\end{equation}
where $\alpha^{~}_h \equiv h^2/(4\pi)$. Using the $90\%$ confidence-level limit
$\Delta a_\mu > 215 \times 10^{-11}$ \cite{Mac}, we then 
obtain $m_\eta < 371 \sqrt{\alpha_h}$ GeV, a mass scale which is not far away
from being discovered or ruled out in the future high-energy collider 
experiments \cite{Ma2}. Furthermore, the branching ratio of 
$l_i \rightarrow l_j \gamma$ is found to be
\begin{equation}
B (l_i \rightarrow l_j \gamma) \; \approx \;
\frac{\alpha}{3072 \pi G^2_{\rm F} m^4_\eta} 
\left | \sum_k (h_{k l_i} h^*_{k l_j}) \right |^2
B(l_i \rightarrow l_j \nu_i \overline{\nu}_j ) \; 
\end{equation}
in the leptonic Higgs doublet model, where $\alpha = 1/137$,
and the relevant assumptions have been taken into account.
For rare decays $\tau \rightarrow \mu \gamma$, $\mu \rightarrow e \gamma$
and $\tau \rightarrow e \gamma$, we obtain
\begin{eqnarray}
\left | \sum_i \left (h_{i\tau} h^*_{i\mu} \right ) \right |^2
& \approx & c^4 h^4 \frac{(\Delta m^2_{\rm atm})^2}{m^4} \;\; ,
\nonumber \\
\left | \sum_i \left (h_{i\mu} h^*_{i e} \right ) \right |^2
& \approx & \frac{c^2}{2} h^4 \left [ \frac{(\Delta m^2_{\rm sun})^2}{m^4} 
+ 4s^2 \frac{(\Delta m^2_{\rm atm})^2}{m^4} \right ] \;\; ,
\nonumber \\
\left | \sum_i \left (h_{i\tau} h^*_{i e} \right ) \right |^2
& \approx & \frac{c^2}{2} h^4 \left [ \frac{(\Delta m^2_{\rm sun})^2}{m^4} 
+ 4 s^2 \frac{(\Delta m^2_{\rm atm})^2}{m^4} \right ] \;\; ,
\end{eqnarray}
in which $\Delta m^2_{\rm sun} \equiv |m^2_2 - m^2_1| \sim 10^{-5} ~ {\rm eV}^2$
and $\Delta m^2_{\rm atm} \equiv |m^2_3 - m^2_1| \approx 
|m^2_3 - m^2_2| \sim 10^{-3} ~ {\rm eV}^2$ are the mass-squared differences
of solar and atmospheric neutrino oscillations \cite{SK}, respectively.
It is clear that $\Gamma (\mu \rightarrow e \gamma)$ or
$\Gamma (\tau \rightarrow e \gamma)$ will get comparable contributions from
the terms associated with $\Delta m^2_{\rm sun}$ and $\Delta m^2_{\rm atm}$,
if $s \sim {\cal O}(10^{-2})$ holds. For larger values of $s$, the
second term may dominate over the first term, leading to the following
interesting relationship:
\begin{eqnarray}
& & \frac{\Gamma (\tau \rightarrow \mu \gamma)}{m^5_\tau} :
\frac{\Gamma (\mu \rightarrow e \gamma)}{m^5_\mu} :
\frac{\Gamma (\tau \rightarrow e \gamma)}{m^5_\tau} 
\nonumber \\
& \approx & c^2 : 2s^2 : 2s^2 \; .
\end{eqnarray}
If the value of $s$ is much lower than ${\cal O}(10^{-3})$, we will arrive at
\begin{eqnarray}
& & \frac{\Gamma (\tau \rightarrow \mu \gamma)}{m^5_\tau} :
\frac{\Gamma (\mu \rightarrow e \gamma)}{m^5_\mu} :
\frac{\Gamma (\tau \rightarrow e \gamma)}{m^5_\tau} 
\nonumber \\
& \approx & 2c^2 (\Delta m^2_{\rm atm})^2 : (\Delta m^2_{\rm sun})^2 : 
(\Delta m^2_{\rm sun})^2 \; .
\end{eqnarray}
This result is consistent with that obtained by Ma and Raidal \cite{Ma}
in the bi-maximal neutrino mixing case with $s=0$.
It becomes transparent that $\Gamma (\mu \rightarrow e \gamma)$ and 
$\Gamma (\tau \rightarrow e \gamma)$ are dependent crucially upon
the magnitude of $s$. In contrast, 
$\Gamma (\tau \rightarrow \mu \gamma)$ is insensitive to small values
of $s$, no matter whether neutrino mixing is bi-maximal or 
nearly bi-maximal.
\begin{figure}[t]
\vspace{-0.83cm}
\epsfig{file=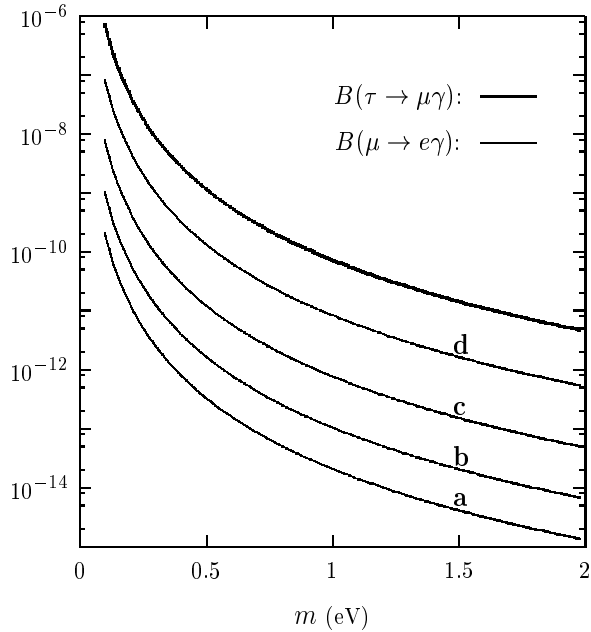,bbllx=-0.5cm,bblly=6cm,bburx=18cm,bbury=28cm,%
width=16cm,height=20cm,angle=0,clip=}
\vspace{-10.5cm}
\caption{The lower bounds on $B(\tau \rightarrow \mu \gamma)$ and
$B(\mu \rightarrow e \gamma)$ in the leptonic Higgs doublet model
with nearly degenerate neutrino masses and nearly bi-maximal neutrino
mixing: (a) $s=0$, (b) $s=0.01$, (c) $s=0.03$, and
(d) $s=0.1$.}
\end{figure}

For illustration, we calculate $B(\mu \rightarrow e \gamma)$ and
$B(\tau \rightarrow \mu \gamma)$ as functions of the neutrino mass
$m$ and the flavor mixing parameter $s$, and plot the numerical
results in Fig. 2, where 
$\Delta m^2_{\rm sun} = 3\times 10^{-5} ~ {\rm eV}^2$,
$\Delta m^2_{\rm atm} = 3\times 10^{-3} ~ {\rm eV}^2$, and
$\alpha_h/m^2_\eta = (371 ~ {\rm GeV})^{-2}$ have been typically
input. The chosen value of $\alpha_h/m^2_\eta$ implies that the 
relevant results shown in Fig. 2 should be understood as {\it lower} 
bounds in the case of nearly bi-maximal neutrino mixing. 
Note that the branching ratio of $\tau \rightarrow e \gamma$ can
straightforwardly be obtained from that of $\mu \rightarrow e \gamma$,
as the relationship 
$B(\tau \rightarrow e \gamma) = B(\mu \rightarrow e \gamma)
\cdot B(\tau \rightarrow e \nu_\tau \overline{\nu}_e)$ with
$B(\tau \rightarrow e \nu_\tau \overline{\nu}_e) \approx 17.83\%$ 
hold \cite{PDG}. From Fig. 2 one can see that 
$B(\mu \rightarrow e \gamma)$ will be close to its current experimental
upper limit of $1.2 \times 10^{-11}$ \cite{ME}, if $m \approx 0.2 ~ {\rm eV}$
and $s \approx 0$ or if $m \approx 1~ {\rm eV}$ and $s \approx 0.1$.
Because of $s\leq 0.2$ \cite{CHOOZ}, no fine tuning is required to
achieve $B(\mu \rightarrow e \gamma) < 1.2 \times 10^{-11}$ 
for $m \leq 2 ~ {\rm eV}$. Neutrinos of $m \sim (1 -2)$ eV may have important
cosmological implications \cite{COS}. To accommodate the nonobservation of the
neutrinoless double beta decay in the scenario of neutrino masses under
discussion, additional Majorana phases should be included into the 
neutrino mixing matrix $V$ \cite{FX00}.

Finally it is worth mentioning that the leptonic Higgs doublet model 
can also lead to the $\mu$-$e$ conversion in nuclei. The relevant assumptions
made above allow us to simplify the formula of the $\mu$-$e$ conversion
ratio $R_{\mu e}$ given in Ref. \cite{Ma}. We then obtain
\begin{equation}
R_{\mu e} \; \approx \; \frac{\alpha^5 m^9_\mu Z^4_{\rm eff} Z
|\overline{F}_p (p_e) |^2}{18432 \pi^4 \Gamma_{\rm capt} q^4 m^4_\eta}
\left | \sum_i \left (h_{i\mu} h^*_{i e} \right ) \right |^2 \; ,
\end{equation}
where $q^2 \approx -m^2_\mu$, $Z_{\rm eff} = 11.62$, $\overline{F}_p = 0.66$, 
and $\Gamma_{\rm capt} = 7.1 \times 10^5 ~ {\rm s}^{-1}$ for 
${\rm Al}$ \cite{MECO}. Comparing between $R_{\mu e}$ and
$B(\mu \rightarrow e \gamma)$, one may easily find that they depend upon 
$m$ and $s$ in the same way. The magnitude of $R_{\mu e}$ is 
smaller than that of $B(\mu \rightarrow e \gamma)$ by two orders. 

In summary, we have discussed the muon g-2 anomaly and the 
lepton-flavor-violating rare decays in the framework of a leptonic
Higgs doublet model with nearly degenerate neutrino masses and 
nearly bi-maximal neutrino mixing. We demonstrate the sensitivity of
$\Gamma (\mu \rightarrow e \gamma)$, 
$\Gamma (\tau \rightarrow e \gamma)$ and $R_{\mu e}$ to possible
non-zero but small values of $|V_{e3}|$. It is obvious that they
are also sensitive to the Dirac-type CP-violating phase in $V$,
which has been taken to be $90^\circ$ in our specific nearly bi-maximal
neutrino mixing pattern. One may of course apply other nearly
bi-maximal neutrino mixing patterns \cite{FXRev} to the models
giving rise to $\Delta a_\mu$ and $\mu \rightarrow e \gamma$.
We expect that new and more precise data to be accumulated from a variety
of experiments on neutrino oscillations and lepton-flavor-violating rare 
decays will help to pin down the self-consistent models (even the true
theory) of neutrino masses, lepton flavor mixing and CP violation.

\vspace{0.5cm}

The author is indebted to X. Calmet for many discussions and partial
involvement at the early stage of this work. He is also grateful
to H. Fritzsch for useful comments and warm hospitality in University
Munich, where this paper was finished.

\end{document}